\documentclass[
reprint, 
superscriptaddress,
showpacs,preprintnumbers,
 amsmath,amssymb,
 aps,
]{revtex4-1}
\usepackage{amsthm}
\usepackage{graphicx}
\usepackage{dcolumn}
\usepackage{bm}
\usepackage{mathrsfs}
\usepackage{dsfont}
\usepackage{color}

\definecolor{cl}{RGB}{255,127,36}
\definecolor{navy}{rgb}{0.0,0.0,0.5}

\begin{document}


\title{A Quantum Complete Prelude to Inflation}

 \author{Stefan Hofmann}%
 \email{stefan.hofmann@physik.uni-muenchen.de}
\affiliation{Arnold Sommerfeld Center for Theoretical Physics, Theresienstra{\ss}e 37, 80333 M\"unchen\\}

\author{Marc Schneider}%
 \email{mms94@psu.edu}
 \affiliation{Institute for Gravitation and the Cosmos, University Park, 16802-6300 Pennsylvania, USA\\}
 
 \author{Maximilian Urban}%
\email{maximilian.urban@physik.uni-muenchen.de}
\affiliation{Arnold Sommerfeld Center for Theoretical Physics, Theresienstra{\ss}e 37, 80333 M\"unchen\\}




\date{\today}

\begin{abstract}
It is shown that the inflationary paradigm admits quantum complete extensions
of  space-time.
The extended inflationary spacetimes still have geodesic borders, 
but quantum fields are prohibited from migrating across these borders by their
evolution semigroups. 
The geodesic singularities lurking across the borders lack a physical description 
because the evolution semigroups give vanishing probabilistic support 
to quantum fields for populating regions bordering on these singularities. 
As an example, anisotropic Bianchi type-I cosmologies are shown to be quantum complete
preludes to inflation. They admit Kasner-like geometries close to their geodesic
borders.
Quantum fields enjoy a contractive evolution in these asymptotic regions
and ultimately become free. As a consequence, quantum probes cannot
migrate across the geodesic border of Bianchi type-I cosmologies. 
\end{abstract}

\pacs{04.20.Dw, 04.62.+v, 11.10.-z, 98.80.Jh}
\keywords{Suggested keywords}
\maketitle


\section{\label{sec:level1}Introduction}
A proper inflationary epoch in the primordial expansion history allows relating
the variety of cosmic structures to quantum fluctuations of the dominant
source during this stage \cite{guth81,linde82,stein82}. These fluctuations represent adiabatic deviations
from local isotropy and constitute the primordial seeds of 
gravitationally driven structure formation. 
As the microscopic origin of large-scale cosmic structures, inflation 
has passed all observational stress tests. 

Inflation extends the cosmic history prior to a decelerated Friedmann cosmology 
with an accelerated expansion stage, which is past-directed timelike and null geodesic 
incomplete and which initially borders on a spacelike singularity.    
However, inflation can emerge from spatially anisotropic but homogeneous cosmologies
as described by Bianchi type-I spacetimes.
These cosmologies develop forward 
instabilities towards approximate isotropic spaces in inflating spacetime regions \cite{Uzan}. 
In addition, Kasner spacetimes develop backward instabilities triggering
Bianchi type-II like transitions \cite{asht}.
between different anisotropic cosmologies, leading to successions of Kasner geometries
\cite{billiards}. 
Kasner spacetimes are past-directed timelike and null geodesic incomplete as well.
They have been shown to be singularity free in canonical quantum cosmology \cite{ash13}.

If geodesic boundaries can be probed in measurement processes, 
they are manifest in observables and imply a physical pathology,
since then observers can be located at the geodesic borders. 
For instance, the singular coincidence limit of Newton's gravitational potential field 
is not sufficient to imply a physical pathology.
A sufficient condition is the existence of initial conditions allowing to realize this coincidence limit dynamically in a finite 
time. Moreover, a repulsive Coulomb potential field is also singular in the coincidence limit,
but since it is unbounded from above near zero, this limit cannot be realized dynamically. The Coulomb potential is 
complete near zero for repulsive charge configurations but incomplete for attractive charge configurations.
However, even for attractive charge configurations, the Coulomb potential 
can still give rise to a complete quantum evolution, as it does for electrons in hydrogen bound states. 

Whether a potential allows for a complete evolution is a context-sensitive question. A similar statement 
does not hold for spacetimes: It is meaningless to conclude that a spacetime supports an
incomplete classical evolution but a complete quantum evolution. If both conclusions are valid, then the 
spacetime is simply incomplete.  The distinction between classical and quantum evolution is still important
because geodesic completeness might fail as a sound physical concept, while it remains a valid 
geometrical characterization of the mathematical spacetime model. 
As a physical concept geodesic completion refers to observations and operational instructions 
for measurements based on classical mechanics. 

In its simplest realization, a measurement process involves three dynamical subsystems: A {\it system} on which 
observers wish to perform measurements using specific {\it apparatuses} in a given {\it spacetime}.
Usually, it is assumed that the inertia of the apparatuses is large compared with that of the system. 
This implies that during the measurement there is no significant backreaction of the system on the 
apparatuses. In addition, it is assumed that the spacetime and its evolution is not affected by the experiment. 
To guarantee the converse, the spacetime region supporting the measurements is required to 
be sufficiently homogeneous. 
Close to the geodesic border of Kasner spacetimes, the latter requirement 
fails to hold for mechanical devices. Tidal forces become too strong during finite time measurements
for these devices to operate as external apparatuses. 
The same is true for recording devices of finite extent with mechanically interconnected
circuit elements. 
 
This leads us to incorporate measurement processes as internal phenomena with no external aspects. 
The system to be examined consists of field configurations and  
the only sensible bookkeeping devices are spectator fields which are locally coupled to these field
configurations. During a measurement all field configurations might be significantly 
altered by spacetime variations. 
The backreaction of all field configurations on spacetime is still required to be negligible.
In this respect, the singularity theorems \cite{hawking}
and our results concerning quantum regularity 
share the same postulate. 
For the sake of keeping the introduction non-technical
this common postulate can be stated informally as follows: The inertia of spacetime is large compared 
to that of all field configurations. In other words, the spacetime is effectively decoupled from 
all field configurations that compose the system and the observers.
This decoupling postulate can be (in)validated a posteriori. 

The ground state of the examined field configurations is time-dependent in dynamical
spacetimes. In general, this time dependence is not given by a phase factor, i.e.~the
evolution of quantum states is not given by a unitary group. One 
can show that even in trivial cases unitarity fails to hold \cite{ag15}.
Here quantum 
states refer to wave functionals over the configuration space in the Schr\"odinger
representation of quantum field theory. 
Instead, the evolution is naturally given
by a semigroup, because the background is assumed to be absolutely inert
against quantum fluctuations.  
A probabilistic interpretation is still possible, provided the semigroup enjoys 
a contraction property that replaces the ordinary isometry requirement. 
Physically the contraction property accommodates damping phenomena 
in the Schr\"odinger Hamiltonian which reduce the normalization of the wave functional. 
This implies that the Schr\"odinger Hamiltonian is not selfadjoint in the physical state space. 
Effectively dynamical spacetimes can act as damping media. 

In contrast to a preinflationary Kasner phase, a pure de Sitter phase
is subjected to a divergent time-evolution towards the initial singularity. The coincidence
limit renders this space-time quantum incomplete. Without 
imposing an initial boundary condition at the time of Big Bang,
finite amplitude tensor fluctuations at finite times will be amplified to trans-Planckian values
when evolved backwards \cite{feld17}. The failure of the no boundary
proposal imposes a general obstruction to de Sitter cosmology close to the 
geodesic border and serves as a plea for a preinflationary phase.

In this article, we show that generic quantum inflaton fields cannot migrate across
the geodesic boundary of Bianchi type-I cosmologies. This is guaranteed by 
a contractive evolution semigroup that assigns a vanishing probability for 
inflaton fields to populate the geodesic border. As a consequence, the spacelike singularity
lurking beyond this border is decoupled from physical spacetime. 
The result is universal in the sense that it holds for any inflaton potential. 
This universality can be interpreted as a consequence of the Belinskii-Khalatnikhov-Lifshitz
conjecture \cite{BKL1,BKL2} which is operative at the geodesic border
and guarantees that the
dynamics at any spatial point is pefectly captured by
 an ordinary differential equation \cite{ash09}.

 \section{Quantum Completeness}
 \label{sec:QC}
Consider a globally hyperbolic spacetime $(\mathcal{M},g)$ and choose a time function $t$
and a vector field $v$ on $\mathcal{M}$ such that the surfaces 
$(\Sigma_t)_{t\in\mathcal{I}\subset\mathbb{R}}$
of constant time are Cauchy hypersurfaces and such that $\nabla_v t=-1$.
Note that the time interval $\mathcal{I}$ might be the entire real line, but we are mostly 
concerned with the time interval $\mathcal{I}:=]0,t_\text{in}]$. We denote by $\mathcal{C}(\Sigma_t)$ the configuration 
space of instantaneous field configurations $\phi$ on $\Sigma_t$, which is infinite-dimensional. 
For simplicity, $\phi:\mathcal{M}\rightarrow\mathbb{K}$, where 
$\mathbb{K}=\{\mathbb{R},\mathbb{C}\}$.  Let $(\mathcal{C}(\Sigma_t),\mathcal{D}\phi)$
denote a formal measure space, and let $\mathcal{L}^2(\mathcal{C}(\Sigma_t))$ denote 
the $\mathbb{C}-$vector space of wave functionals $\Psi_t: \mathcal{C}(\Sigma_t)\rightarrow \mathbb{C}$
which are measureable and whose modulus is square integrable with respect to the functional measure $\mathcal{D}\phi$.
We introduce the usual seminorm on $\mathcal{L}^2(\mathcal{C}(\Sigma_t))$: 
\begin{eqnarray}
	\Vert\Psi_t\Vert_2
	:=
	\left\{
		\int_{\mathcal{C}(\Sigma_t)}
		\mathcal{D}\phi \; \left|\Psi_t[\phi]\right|^2 
	\right\}^{1/2}	
\end{eqnarray}
In general this is not a norm on $\mathcal{L}^2(\mathcal{C}(\Sigma_t))$ since
$\Vert\Psi_t\Vert_2=0$ only implies $\Psi_t[\phi]=0$ almost everywhere in $\mathcal{C}(\Sigma_t)$
but not $\Psi_t\equiv 0$. For this reason we introduce
\begin{eqnarray}
	\mathcal{N}(\mathcal{C}(\Sigma_t),\mathcal{D}\phi)
	&:=&
	\left\{\right.
		\Psi_t\in\mathcal{L}^2(\mathcal{C}(\Sigma_t),\mathcal{D}\phi)
		: 
		\Psi_t[\phi] =0 
		\nonumber \\
		 &&\mathcal{D}\phi-\text{almost everywhere}
	\left.\right\}
	\; .
\end{eqnarray}
$\mathcal{N}(\mathcal{C}(\Sigma_t),\mathcal{D}\phi)$ is a subspace of 
$\mathcal{L}(\mathcal{C}(\Sigma_t),\mathcal{D}\phi)$, and $\Vert\cdot\Vert_2$ 
is a norm on the quotient space 
\begin{eqnarray}
	L^2 (\mathcal{C}(\Sigma_t),\mathcal{D}\phi)
	:=
	\mathcal{L}^2(\mathcal{C}(\Sigma_t),\mathcal{D}\phi)
	\setminus \mathcal{N}(\mathcal{C}(\Sigma_t),\mathcal{D}\phi)	\; .	
\end{eqnarray}
Note that $L^2 (\mathcal{C}(\Sigma_t),\mathcal{D}\phi) $ is not a space of 
wave functionals, but a space of equivalence classes of wave functionals. 
All operations are defined with respect to representatives.

Concerning interpretation, $|\Psi_t[\phi]|^2$ is a probability density in 
the following sense: If $\mathcal{U}$ is a (measurable) subset of $\mathcal{C}(\Sigma_t)$
and $\mathcal{X}_\mathcal{U}$ its indicator functional, then $\Vert\mathcal{X}_\mathcal{U} \Psi_t\Vert_2^{\; 2}$
is the probability for the field configuration on $\Sigma_t$ to be given by some $\phi\in\mathcal{U}$.
This interpretation requires $\Vert\Psi_t[\phi]\Vert_2=1$ or at least $\Psi_t$ to be a normalizable functional
for $t\in\mathcal{I}$. The smeared configuration field operator $\Phi[f]$ 
is just the operator for multiplication with $\phi[f]$, where $f$ is a smooth smearing function of compact 
support in $\Sigma_t$ such that the expectation value 
$\langle\Psi_t|\Phi[f]|\Psi_t\rangle=\Vert \sqrt{|\phi[f]|}\Psi_t\Vert_2^{\; 2}$ is well defined.
As usual a measure for the scatter of field configurations around the expectation value 
is given by $\Vert (\Phi[f]-\phi[f]\text{id}_{L^2})\Psi_t\Vert_2$.
The momentum field operator $\Pi[f]$
is the functional derivative $-\text{i}(\text{det}(g_{_{\Sigma_t}}))^{-1/2}\delta/\delta\phi$
in the direction of a smooth function $f$ of compact support. 
Here $g_{_{\Sigma_t}}$ denotes the pullback of the metric field $g$ to the hypersurface $\Sigma_t \; (t\in\mathcal{I})$.
Heisenberg's fundamental uncertainty relation follows from 
$[\Phi[f_1],\Pi[f_2]]\Psi_t[\phi]=\text{i}\langle f_1,f_2\rangle \Psi_t[\phi]$,
where $\langle\cdot,\cdot\rangle$ denotes the canonical scalar product
for smearing functions on $\Sigma_t$. It should be stressed that the 
above construction is only meant at a formal level.

We now consider a family of formal evolution operators 
$\{\mathcal{E}_g(t,t_0): 0\leq |t-t_0|<\infty\}$ on 
$L^2(\mathcal{C}(\Sigma_t),\mathcal{D}\phi)$. Such a family is a strongly 
continuous semigroup if $\mathcal{E}_g(t_0,t_0) = \text{id}_{L^2}$,
$\mathcal{E}_g(s,t) \mathcal{E}_g(t,t_0) = \mathcal{E}_g(s,t_0)$ for all 
$s,t,t_0\in ]0,\infty[$ with $s,t\leq t_0$ (which reflects our convention 
that the flow of time runs from $t_0$ towards $0$), and 
if for each $\Psi_t\in L^2(\mathcal{C}(\Sigma_t),\mathcal{D}\phi)$ 
the map $\mathcal{I}\subset\mathbb{R}\rightarrow L^2(\mathcal{C}(\Sigma_t),\mathcal{D}\phi)$, 
defined by $t\mapsto \mathcal{E}_g(t,t_0)\Psi$ is continuous. 
Such evolution semigroups arise naturally in the Schr\"odinger-picture quantum
theory of fields in curved spacetimes. A probabilistic interpretation is only possible for 
a special class of evolution semigroups: A contractive evolution semigroup 
satisfies the additional requirement $\Vert \mathcal{E}_g(t,t_0)\Vert \leq 1$
for all $t\leq t_0$ in the time interval $\mathcal{I}$. Here the operator norm is 
defined as usual, $\Vert \mathcal{E}_g(t,t_0)\Vert := 
\text{inf}\{C\geq 0 : \Vert \mathcal{E}_g(t,t_0)\Psi\Vert_2 \leq C \Vert\Psi\Vert_2 \; 
\text{for all}\;  \Psi_t\in L^2(\mathcal{C}(\Sigma_t),\mathcal{D}\phi) \}$.

As in the case of unitary evolution groups we obtain the generator of $\mathcal{E}_g(t,t_0)$
by differentiation with respect to $t$. Set
$h_g(t):= (\text{id}_{L^2}-\mathcal{E}_g(t,t_0))/|t-t_0|$ and consider only 
$\Psi_t\in L^2(\mathcal{C}(\Sigma_t),\mathcal{D}\phi)$ for which 
$h_g(t)\Psi_t$ exists in the limit $|t-t_0|\rightarrow 0$. We denote this limit by $h_g(t_0)\Psi_{t_0}$
and call $h_g(t_0)$ the infinitesimal generator of $\mathcal{E}_g(t,t_0)$.
Of course, $h_g = H[\Phi,\Pi;g]$, where $H$ denotes the Hamiltonian 
composed of configuration field operators $\Phi$ and conjugated momentum field 
operators $\Pi$ in $L^2(\mathcal{C}(\Sigma_t),\mathcal{D}\phi)$. 
For simplicity we focus on a simple local quantum theory with a Hamiltonian density 
$\mathcal{H}=\mathcal{T}(\Pi;g)+\mathcal{V}(\Phi;g)$. On the hypersurface $\Sigma_t$
the functional Laplacian is 
$\mathcal{T}:=\sqrt{g_{tt}}\, \Pi\circ\Pi/2$, where 
$\mathcal{V}$ denotes the 
effective potential with $\sqrt{g_{tt}} \, g_{_{\Sigma_t}}^{-1}(\text{d}\Phi,\text{d}\Phi)$ included.
The infinitesimal generator $h_g$ and the evolution operator $\mathcal{E}_g(t,t_0)$ 
are related as follows:
\begin{eqnarray}
	&&\mathcal{E}_g(t,t_0)
	=
	T_{\leftarrow}\exp{
		\left\{
			-\text{i} \int_{t_0}^t {\rm d}t^\prime H[\Phi,\Pi;g](t^\prime)
			\right\}}
			\\
	 &&H[\Phi,\Pi;g](t^\prime)
	=
	\int_{\Sigma_{t^\prime}}{\rm d}\mu \;\mathcal{H}(\Phi,\Pi;g_{_{\Sigma_{t^\prime}}}) 
	\; .
\end{eqnarray}
Here $\text{d}\mu$ is the covariant measure on $\Sigma$ and $T_{\leftarrow}$ denotes the time-ordering
starting at $t_0 >t$.

We now turn to the crucial condition for the generator of an evolution semigroup. 
Consider a dual element $S\in [L^2(\mathcal{C}(\Sigma_t),\mathcal{D}\phi)]^*$ which
satisfies $\Vert S\Vert=\Vert\Psi\Vert_2$ and $S(\Psi_t)=\Vert\Psi_t\Vert_2^{\; 2}$.
We can think of $S$ as a normalized tangent functional to $\Psi_t$. The Hahn-Banach
theorem guarantees that each wave functional $\Psi_t\in L^2(\mathcal{C}(\Sigma_t),\mathcal{D}\phi)$
has a normalized tangent functional. The generator $h_g$ of $\mathcal{E}(t,t_0)$ is called 
accretive if for each $\Psi_t\in L^2(\mathcal{C}(\Sigma_t),\mathcal{D}\phi)$ we have
$\text{Im}(S(h_g\Psi_t))\leq 0$. The relation between contractive and accretive is 
almost straight forward: $h_g$ is the generator of a contractive evolution semigroup if and only if
$h_g$ is accretive.

Suppose $\Sigma_0$ is a spacelike geodesic boundary of $(\mathcal{M},g)$ located at $t\rightarrow 0$. 
We call $(\mathcal{M},g)$ quantum complete if the 
evolution semigroup of wave functionals on test-field configurations  
is contractive in $(\mathcal{M},g)$ and 
$\Vert\mathcal{E}(t,t_0)\Vert\rightarrow 0$ for $t\rightarrow 0$ \cite{hof15}. 
This definition of quantum completeness tacitly assumes that the test-field configurations
enjoy a healthy evolution in Minkowski spacetime.

\section{Anisotropic Prelude}
\label{sec:AP}
The prelude to inflation considered in the work is given by an anisotropic Bianchi type-I
cosmology with a sequence of Kasner spacetimes characterizing the neighborhood 
of its geodesic boundary.  A Kasner-like preinflationary
phase is consistent with observational bounds on the e-folds because
of the fast developing forward instability \cite{Kofman}. 
Any spacetime $(\mathcal{M},g)$ considered in this prelude is a 
multiple warped product manifold of the form $\mathcal{M}= 
\mathcal{I}\times_{w_1}\mathbb{R}\times_{w_2} \mathbb{R}\times_{w_3} 
\mathbb{R}$
furnished with a tensor field
\begin{eqnarray}
	\label{defWP}
	g = -\pi^*\left(\mbox{d}t\otimes\mbox{d}t\right) +
	\sum_{a=1}^3 \left(w_a\circ\tau\right)^2 \sigma_a^{\; *} \left(q_a\right)
	\; ,
\end{eqnarray}
with positive warping functions $w_a\in C^\infty(\mathcal{I}) \; (a\in\{1,2,3\})$.
By $\tau$ and $\sigma_a$ we denote the projections onto the base (time interval) $\mathcal{I}$
and the fibers (in our case one-dimensional subspaces), respectively, and $\pi^*$ and $\sigma_a^{\; *}$ are 
the correponding pullbacks. The pairs $(\mathbb{R},q_a)$ denote the flat Riemannian 
Fiber manifolds with respect to the base manifold 
$(\mathcal{I}, -\mbox{d}t\otimes\mbox{d}t)$. 
Kasner spacetimes are multiple warped products of this type with 
warping functions $w_a = \text{id}_\mathcal{I}^{\; p_a}$, where $p_a\in\mathbb{R}$ denote 
the so-called Kasner exponents. The Kasner exponents are required to lie in the 
intersection of the Kasner plane $p_1+p_2+p_3=1$ and the Kasner sphere 
$p_1^{\; 2}+p_2^{\; 2}+p_3^{\; 2}=1$.
Kasner spacetimes can be characterized as Ricci-flat Einstein manifolds which 
are globally hyperbolic, future-directed time-like and null geodesic complete, but past-directed time-like and null 
geodesic incomplete. Since they are vacuum solutions, Kasner geometries can only be an approximate
description close to the geodesic border of an anisotropic spacetime such as the more general Bianchi type-I models
relevant for this work. Bianchi type-I geometries are multiple warped product manifolds $\mathcal{M}_B$
of the type (\ref{defWP}).

Consider a classical scalar field $\phi:\mathcal{M}_B\rightarrow \mathbb{R}$ in a Bianchi type-I spacetime with
Hamilton density $\mathcal{H}=\mathcal{T}(\pi; g) + \mathcal{V}(\phi;g)$. The effective 
potential $\mathcal{V}$ includes the inflaton potential \cite{Gumruckzuck1}
\begin{eqnarray}
	\label{IP}
	V=V_{\rm dS}\left(1-\exp{\left(-\phi/\phi_0\right)}\right)^2
	\; .
\end{eqnarray}
This potential is harmonic in $\phi$ around its minimum and dominated 
by a spacetime homogeneous energy density $V_{\rm dS}$ away from it. 
The potential parameters are fixed as $\sqrt{G_{\rm N}} \phi_0=10^{-3}$ and
$V_{\rm dS}=10^{13}\,$GeV in order to obtain the correct amplitude of metric 
perturbations in the $SO(3)-$scalar sector. 

In the slow-roll regime $\mathcal{T}\ll V_{\rm dS}$ at early times $t\ll 1/H_{\rm dS}$
the anisotropic expansion history is given by a Kasner solution with warping 
functions $w_a(t)/w_a(t_{\rm in}) \approx (t/t_{\rm in})^{p_a}$ for $a\in\{1,2,3\}$.
At these early times the approximate de Sitter source 
$H_{\rm dS}^{\; 2}=(8\pi/3) G_{\rm N} (V_{\rm dS}+\mathcal{O}(\mathcal{K}/V_{\rm dS}))$ 
is effectively decoupled from spacetime and Kasner geometries can emerge.
At intermediate times $t\sim 1/H_{\rm dS}$ the expansion history is given by a 
more general Bianchi type-I solution of Einstein's equation \cite{Gumruckzuck2},
$w_a(t)/\overline{w_a}=[\sinh{(3 t H_{\rm dS})}]^{1/3} [\tanh{(3 t H_{\rm dS}/2)}]^{p_a-1/3}$,
where $\overline{w_a}$ denote normalization constants. While the exponents 
$p_a$ are in the intersection of the Kasner plane and Kasner sphere, the warping functions 
explicitly depend on the transition time $1/H_{\rm dS}$.  
For late times $t\gg 1/H_{\rm dS}$ an approximate de Sitter stage emerges with 
$w_a(t)/w_a(t_0)\approx\exp{((t-t_0) H_{\rm dS})}$. 

Clearly $1/H_{\rm dS}$ is the characteristic time scale for isotropization driven by $V_{\rm dS}$:
Consider the Weyl tensor as a tensor of type $(0,4)$ given by 
$C=R-\tfrac{1}{2}(\text{Ric}-\tfrac{1}{4} S \, g)\ast g - \tfrac{1}{24} S \, g\ast g$, where 
$R$ denotes the type-$(0,4)$ Riemann tensor, Ric is the Ricci tensor and $S$ the 
curvature scalar. For any symmetric tensors $T_1,T_2$ of type $(0,2)$,  
$T_1\ast T_2$ denotes the Kulkarni-Nomizu product.
Throughout the evolution up to the end of inflation $\text{Ric}\approx\tfrac{1}{4} S\, g$, so 
$C\approx R-\tfrac{1}{24} S\, g\ast g$. The Ricci decomposition is therefore 
$|R|^2 \approx |C|^2 + |\tfrac{1}{24} S\, g\ast g|^2$. At early times $t\ll 1/H_{\rm dS}$
this decomposition into irreducible components with respect to the orthogonal group
is dominated by the anisotropic contribution, $|R|^2 \approx |C|^2=|p_1 p_2 p_3| (2/t)^4$.
The asymptotic behavior of the conformal tensor renders any de Sitter-like source  
initially irrelevant, which is why the Kasner solution is a good description of the geometry
in the vicinity of the cosmic singularity. In contrast, at late times $t\gg 1/H_{\rm dS}$
the Ricci decomposition is dominated by the approximate de Sitter source
$|R|^2\approx |\tfrac{1}{24} S\, g\ast g|^2$ because $|C|^2\approx \exp{(-6 t H_{\rm dS})}$.
In this stage, the approximate de Sitter source damps all anisotropic contributions to the 
curvature. 

\section{Kasner spacetimes are quantum complete}
The last section was devoted to geometric preliminaries of anisotropic 
preludes to inflation. These are given by approximate Kasner spacetimes 
which exhibit a forward instability towards more general Bianchi type-I
cosmologies that allow for isotropization at a later stage.
Kasner spacetimes border on singular hypersurfaces $\Sigma_0$ as well. 
This raises the question of how their geodesic boundary compares
to the geodesic incompleteness of isotropic cosmologies.
At the intersection of the Kasner plane and sphere 
$p_a\in[-1/3,1[$, excluding the case $p_a=0$ for some $a\in\{1,2,3\}$
which is isometric to an accelerated frame in Minkowski spacetime.
The most interesting case is given by
$p_a=-1/3$ for some Kasner index, then the remaining two exponents 
are equal to $2/3$. For definiteness and without loss of generality 
consider $p_1=-1/3, p_2=p_3=2/3$. This is the unique solution with 
$p_2=p_3$. In general, the above half-closed interval for the Kasner 
exponents forces us to restrict the base (time interval) to 
$\mathcal{I}=]0,t_{\rm in}]$, where for the upper limit $t_\text{in}\ll 1/H_{\rm dS}$.  
Of course, for the exact Bianchi type-I solution $\mathcal{I}=]0,\infty[$, 
but our focus is on the vicinity of the geodesic border $\Sigma_0$. 
Consider two spacetime events $P=(d_{\rm in}, 0, 0)$ and $Q=(0,0,0)$
relative to some coordinate neighbourhood in 
the hypersurface $\Sigma_\text{in}$. Towards the geodesic 
border $\Sigma_0$ the proper spatial distance $d_{\rm phys}$
between these events scales as 
$d_\text{phys}(t)/d_\text{in}=(t_\text{in}/t)^{2/3}$. 
Coincidence limits of events towards the singularity 
are prohibited by the Kasner conditions, while Friedmann 
cosmologies enforce coincidence limits, no matter 
how they are regularized in interacting field theories. 
This observation gives rise to the hypothesis that 
Kasner spacetimes are quantum complete in the sense of 
section \ref{sec:QC}. 

Consider the space $\mathcal{C}(\Sigma_t)\, (t\in\mathcal{I})$ of instantaneous
inflaton field configurations on the hypersurface $\Sigma_t$. We introduce a
bilinear functional $\mathcal{K}_t:\mathcal{C}(\Sigma_t)\times\mathcal{C}(\Sigma_t)\rightarrow \mathbb{C}$,
$(\phi_1,\phi_2)\rightarrow [\phi_1] \mathcal{K}_t[\phi_2]$ with coordinate representation
\begin{eqnarray}
	[\phi_1]\mathcal{K}_t[\phi_2]
	=
	\int_{\Sigma_t} \text{d}\mu(x) \text{d}\mu(y) \, \phi_1(x) K_t(x,y) \phi_2 (y)\, ,
\end{eqnarray}
where $\text{d}\mu(x)=\sqrt{\text{det}(q)} \text{d}^3x$ and $q(t)$ denotes the spatial metric field induced on $\Sigma_t$.
The bilocal kernel function $K$ is spatially homogeneous and
the bilinear functional $\mathcal{K}$ is symmetric. In addition to spatial homogeneity the scaling of
$d_\text{phys}(t)/d_\text{in}=(t_\text{in}/t)^{2/3}$ implies that initially separated 
events remain uncorrelated close to the geodesic border, as is shown below. Events can only be correlated if 
the spacetime regions supporting them intersect. Before calculating the kernel, we may therefore guess that
$K(x,y;t)=k(t) \delta(x-y)/\sqrt{\text{det}(q)(t)}$ in the vicinity of the geodesic singularity. It follows that
\begin{eqnarray}
	[\phi_1]\mathcal{K}_t[\phi_2]
	\rightarrow	
	k(t) \int_{\Sigma_t} \text{d}\mu(x) \, \phi_1(x) \phi_2(x) 
\end{eqnarray}
for $t\rightarrow 0$.
If we specialize to $[\phi]\mathcal{K}_t[\phi]$ then the bilinear functional 
becomes a quadratic functional $\mathcal{K}_{t} [\phi]$ in this limit.
Since the inflaton field experiences exclusively ultra-local 
self-correlations, a regularization prescription is required. 
Note that towards the geodesic border $\mathcal{K}_{t}[\phi]$
is similar to a time-dependent mass term in the Lagrange function for the inflaton field.


In the following we consider smooth functions on $\Sigma_t$ and linear functionals on the algebra
of these smooth functions, but our notation will not distinguish between them. 
The bilinear functional $\mathcal{K}_t$ can be described by on-shell configuration fields in
$\mathcal{S}=\{\varphi:\mathcal{I}\times\Sigma_{t\in\mathcal{I}}\rightarrow \mathbb{C}: (\Box-m^2)\varphi=0\}$.
Huygens' principle relates the kernel functional with on-shell configuration fields as follows 
\begin{eqnarray}
\label{Huygens}
	J\left[-\text{i} \partial_t\varphi\right]
	= 	
	[J]\mathcal{K}_{t}[\varphi]
\end{eqnarray}
for any smooth detector source $J$.
A more direct relation holds in Fourier space. Our convention for the Fourier transform is
\begin{eqnarray}
	\mathcal{F}_t f(k) :=
	\int_{\Sigma_t}\text{d}^3x \, \exp{\left(-\text{i}k\cdot x\right)} f(t,x)
\end{eqnarray}
for any $f$ in $L^1(\Sigma_t)$ or Schwartz space. The argument of the plane wave is
$k\cdot x := k_a x^a$. For all $f$ in Schwartz space, the function $f$ can be recovered from its Fourier
transform by the inversion formula
\begin{eqnarray}
	\mathcal{F}^{-1} \widehat{f} (x) = (2\pi)^{-3} \int_{\mathbb{R}^3}
	\text{d}^3 k \, \exp{\left(\text{i}k\cdot x\right)} f(t,k)
	\; ,
\end{eqnarray}
where $\text{d}^3 k := {\text d}k_1 \wedge  {\text d}k_2 \wedge {\text d}k_3$.
For smooth fields (\ref{Huygens}) implies
$-\text{i}\partial_t\varphi=K_t\star\varphi$, 
with $K_t\star\varphi$ denoting the covariant convolution of the bilocal kernel function
and the configuration field on $\Sigma_t$.
Using 
$\mathcal{F}^{-1}\widehat{K_t} \star \mathcal{F}^{-1}\widehat{\varphi}=
\mathcal{F}^{-1}\widehat{K_t} \widehat{\varphi}$, where the Fourier transforms 
are with respect to the convolution variable, and 
$\partial_t\mathcal{F}^{-1}\widehat\varphi = \mathcal{F}^{-1}\partial_t\widehat{\varphi}$, 
Huygens's principle (\ref{Huygens}) can be written as
\begin{eqnarray}
	\frac{-\text{i}}{\sqrt{\text{det}(q)}} \, \partial_t \, \text{ln} \, \frac{\widehat{\varphi}(t,k)}{\widehat{\varphi}(\tau,k)}
	=
	\widehat{K}(t,k) \; ,
\end{eqnarray}
where $\tau$ is an arbitrary reference time in $\mathcal{I}=]0,t_{\rm in}]$. 

%
On-shell field configurations are easily calculated in the vicinity of the geodesic border 
to Kasner spacetimes. The d'Alembert operator is 
$\Box = \partial_t^{\, 2} + t^{-1} \partial_t - t^{-2p_a}\partial_a^{\, 2}$ with the spatial 
index running in $a\in\{1,2,3\}$. The ansatz $\varphi(t,x)=T(t) R(x)$
gives an ordninary second-order differential equation for $T$ with singular coefficients:
$(\partial_t^{\, 2}+t^{-1}\partial_t -\kappa^\prime t^{-2\alpha} \, {\rm id})T\approx 0$. 
Here $\alpha$ is the largest Kasner exponent, $\kappa^\prime$ is determined from the 
solution of $R$ and $\approx$ denotes equality up to
irrelevant contributions, i.e.~contributions which are less singular than those given.
We introduce a pivotal time scale $t_*$ and $\varepsilon \tau:=t/t_*$,
where $\varepsilon >0$ is a smallness parameter. Then
$(\partial_\tau^{\; 2} - \tau^{-1}\partial_\tau) T\approx +\kappa \varepsilon^{2(1-\alpha)} \tau^{-2\alpha} T$. 
Since $p_a\in[-1/3,1[$ the largest Kasner exponent is smaller than one. In the limit 
$\varepsilon\rightarrow 0$ the asymptotic solution is $T\approx c_1 + c_2 \text{ln}(\varepsilon\tau)$
with $c_1,c_2\in\mathbb{R}$. If the Kasner spacetime is axisymmetric then explicit 
solutions are known in terms of Bessel and biconfluent Heun functions. Close to 
the geodesic border these agree 
with our asymptotic solution $\varphi(t,x)=C_1(x) + C_2(x) \text{ln}(t/t_*)$, where 
$C_i \, (i\in\{1,2\})$ are smooth complex valued functions on $\Sigma_t$ and $t,t_*\in\mathcal{I}$.
For simplicity we choose $t_*=t_\text{in}$ for the pivotal time scale such that
$\varphi(t_\text{in},x)=C_1(x)\;, \dot{\varphi}(t_\text{in},x)=C_2(x)/t_\text{in}$ and 
$\varphi(t,x)=\varphi(t_\text{in},x) - t_\text{in}\dot{\varphi}(t_\text{in},x) \mid \text{ln}(t/t_\text{in})\mid$
for $t\in]0,t_\text{in}]$. As a result the kernel function is given by
\begin{eqnarray}
	\label{solution}
	K_t(x,y)
	\approx
	- \frac{\text{i}}{t^2 \text{ln}\left(t/t_{\rm in}\right)}
	\left[\delta\left(x-y\right)-\frac{R(x,y)}{\text{ln}\left(t/t_{\rm in}\right)}\right] 
	\, .
\end{eqnarray}
Here $R:=\mathcal{F}^{-1} \widehat{C_1}/\widehat{C_2}$.
To leading order in $\text{ln}^{-1}(t/t_{\rm in})$ $(t\in]0,t_{\rm in}[)$ the kernel
function is given by a purely imaginary contact term. This contribution alone is a particular solution 
to Huygens's principle (\ref{Huygens}). It is universal in the following sense: Let us introduce a functional 
generalization of the Dirac measure 
$\delta: \mathcal{C}_0^\infty(\Sigma_t)\times \mathcal{C}_0^\infty(\Sigma_t)\rightarrow \mathbb{C}$,
defined by $[f_1]\delta[f_2] := f_1[f_2]$. Note that our notation does not distinguish
between elements of $\mathcal{C}_0^\infty(\Sigma_t)$ and linear functionals on $\mathcal{C}_0^\infty(\Sigma_t)$.
A particular kernel functional solving (\ref{Huygens}) is given by
$[f_1]\mathcal{K}_t[f_2]
=
-\text{i} [f_1]\delta[f_2\, \partial_t  \text{ln}\left(f_2/f_{2{\rm in}}\right)]$ can be extended 
to a bilinear functional associated with propagating waves. On its own it accounts only for the
wave front at the location of the detector described by the current density $J$, but does not contain 
information about spatial correlations between wave fronts on $\Sigma_t$, which is 
partially contained in the bilinear functional $\mathcal{R}$ associated with $R$. 
This contribution, however, is already subleading in the vicinity of 
the geodesic border. Hence events at different spatial locations on $\Sigma_t$ are 
approximately uncorrelated close to the geodesic border. In other words, towards the geodesic singularity 
the spacetime probes consisting of quantum fields are reduced to decoupled pointlike degrees of freedom. 
The latter statement will be substantiated in the remainder of the article.

Close to the classical equilibrium configuration of the inflaton potential (\ref{IP}) we make the 
following ansatz for the Schr\"odinger wave functional:
\begin{eqnarray}
\label{ansatz}
\Psi_t[\phi] = \Psi_t^{(0)}[\phi] \times \exp{\left(\mathcal{D}_t[\phi]\right)} \; .
\end{eqnarray}
Here $\Psi_t^{(0)}$ denotes the ground state functional 
\begin{eqnarray}
	&&\Psi_t^{(0)}[\phi]
	=
	\mathcal{N}_t^{(0)} \mathcal{G}_t^{(0)}[\phi]
	\; ,
	\nonumber \\
	&&\mathcal{N}_t^{(0)}
	=
	\mathcal{N}_{t_\text{in}}^{(0)}
	\exp{
	\left\{
		+\tfrac{\rm i}{2}
		\int_{t_{\rm in}}^t \text{d}\tau \int_{\Sigma_\tau}\text{d}\mu \, \tfrac{1}{2}\Pi^2 \, [\phi]\mathcal{K}_\tau[\phi]
	\right\}
	}
	\; ,
	\nonumber \\
	&&
	\mathcal{G}_t^{(0)}[\phi]
	=
	\exp{
	\left\{
		-\tfrac{1}{2} [\phi]\mathcal{K}_t[\phi]
	\right\}
	}
	\; ,
\end{eqnarray}
and $\mathcal{D}_t[\phi]$ generates non-Gaussian deformations of the ground state 
due to inflaton self-interactions caused by the potential (\ref{IP}).
Note that $\mathcal{D}_t[\phi] = \sum_{n\ge 2} \mathcal{D}^{[n]}_{t}[\phi]$
is a sum of nonlinear functionals starting at quadratic order. 

Consider the wave functional of the ground state in the vicinity of 
the geodesic border $\Sigma_0$. 
To leading and subleading order 
its normalization is given by
\begin{eqnarray}
	&&\frac{\mathcal{N}^{(0)}_t}
	{\mathcal{N}^{(0)}_{t_{\rm in}}}
	\approx 
	\exp{\left\{
		-\tfrac{1}{2} \text{vol}_{\rm ps} \int_{t_\text{in}}^t \frac{\text{d}\tau}{\tau\text{ln}(\tau/t_\text{in})}
	\right\}}
	\times 
	\nonumber\\
	 &&\times
	 \exp{\left\{
		+\tfrac{1}{2} R(0) \text{vol}\left(\Sigma_t\right) \int_{t_\text{in}}^t \frac{\text{d}\tau}{\tau\text{ln}^2(\tau/t_\text{in})}
	\right\}}
	\;, 
\end{eqnarray}
where $\text{vol}_{\rm ps}=\text{vol}(\Sigma_t) \, \text{vol}(T^*\Sigma_t)$ denotes the time-independent coordinate 
phase space volume, and $R(0)$ is the regularized value of the complex-valued bilocal function $R$ in the coincidence limit. 
As anticipated the kernel function requires regularisation in the spatial coincidence limit
on any hypersurface $\Sigma_t \, (t\in]0,t_\text{in}])$. In particular, since $\text{vol}_\text{ps}$ is time-independent
this requirement is logically independent from the existence of a geodesic border. 
Therefore it is sufficient to introduce crude cut-off regulators for the purposes of this article.
We find 
\begin{eqnarray}
	&&\left[\text{ln}\left(\mathcal{N}^{(0)}_\tau\right)\right]_{t_{\rm in}}^t
	\approx
	-\frac{\text{vol}_{\rm ps}}{2} 
	\left[
		\text{ln}\left(\text{ln}\left(\tau/t_{\rm in}\right)\right)
	\right] _{t_{\rm in}}^t
	\nonumber \\
	&&-\frac{\text{vol}(\Sigma_t)}{2} R(0)
	\left[
		\text{ln}^{-1}\left(\tau/t_{\rm in}\right)
	\right] _{t_{\rm in}}^t
	\; .
\end{eqnarray}
For $t/t_\text{in}\rightarrow 0$ the ground state normalization $\mathcal{N}^{(0)}_t\propto \exp{(\Gamma_t)}$ is
exponentially suppressed with a damping factor $\Gamma_t \approx (-\text{vol}_{\rm ps}/2)\text{ln}(|\text{ln}(t/t_\text{in})|)$
up to an irrelevant phase in leading order. The real
part of the damping factor monotonically approaches
minus infinity when $t$ goes to zero.

In order to study $\mathcal{G}^{(0)}$ we decompose the bilocal kernel function $R=\text{Re}(R)+\text{i}\,\text{Im}(R)$
and its associated bilinear functional $\mathcal{R}=\mathcal{R}_\text{re}+\text{i}\mathcal{R}_\text{im}$
accordingly. Then 
\begin{eqnarray}
	\label{G0}
	\mathcal{G}^{(0)}_t
	&\approx &
	\exp{
		\left\{
			\frac{\text{i}}{2}\frac{1}{t^2\text{ln}(t/t_\text{in})} [\phi]\left(\delta-\frac{\mathcal{R}_\text{re}}{\text{ln}(t/t_\text{in})}\right)[\phi]
		\right\}
	}	
	\nonumber \\
	&\times &\exp{
		\left\{
			\frac{1}{2}\frac{1}{t^2\text{ln}^2(t/t_\text{in})} [\phi]\mathcal{R}_\text{im}[\phi]
		\right\}
	}
\end{eqnarray}
towards the geodesic border $\Sigma_0$. Note that for any bilinear functional 
$\mathcal{A}\in\{\delta,\mathcal{R}_\text{re},\mathcal{R}_\text{im}\}$
appearing in (\ref{G0}) the combination $t^{-2}\mathcal{A}$ is time independent. Therefore
$\mathcal{G}^{(0)}[\phi]\rightarrow 1$ in the limit $t/t_\text{in}\rightarrow 0$ for any 
field configuration. 
 
As a result $\Psi_t^{(0)}[\phi]\rightarrow 0$ towards $\Sigma_0$, which implies 
\begin{eqnarray}
\lim_{t\rightarrow 0} \left\Vert\Psi_t^{(0)}\right\Vert_2=0 \; .
\end{eqnarray}
The ground state wave functional does not yield probabilistic support to any field configuration 
at the geodesic border. 
While Kasner spacetimes border at a geodesic singularity, they do not leak information 
across the geodesic border. There is no physical characterization of $\Sigma_0$ in terms 
of measurement processes and observables. 
By direct Kernel methods we have 
shown that $\{\mathcal{E}_g(t,t_\text{in}): t\in]0,t_\text{in}]\}$ is a contraction 
semigroup describing the ground state evolution in asymptotic Kasner spacetimes. 
Since $\mathcal{E}_g(t,t_\text{in})$ has an explicit kernel, we can qualify asymptotic Kasner 
geometries as quantum complete preludes to inflation (via intermediary Bianchi type-I cosmologies)
by direct kernel methods. 
The succession from asymptotic Kasner geometries via Bianchi type-I to inflationary spacetimes 
is, therefore, a quantum complete sequence of physical spacetimes which is
consistent with the results in \cite{giel16}. The singular potential in anisotropic
cosmologies has no effect on the consistency of scattering processes. 
In turn, extending inflation by an anisotropic prelude with asymptotic Kasner geometry 
results in a quantum complete inflationary paradigm. 

Non-Gaussian deformations of the ground state are generated by self-interactions. We refrain from presenting the 
calculation of $\mathcal{D}_t$ here. The basic result is that self-interactions do not alter the
leading asymptotic behavior towards the geodesic border. 
Intuitively this can be understood 
by direct kernel methods \cite{hof17} as follows. 
The Hamilton density operator of the inflaton is $\mathcal{H}=\mathcal{T}(\Pi;g) + \mathcal{V}(\Phi;g)$,
where $\mathcal{T}=(\Pi)^2/2=-1/(2\, \text{det}(q))\delta^2/\delta\phi^2$ denotes
the functional Laplacian and $\mathcal{V}=q^{-1}(\text{d}\Phi,\text{d}\Phi) + V(\Phi)$ 
the effective potential density operator including the inflaton potential 
$V=V_{\rm dS} \, (\Phi/\phi_0)^2+V_{\rm dS} \, \mathcal{O}[(\Phi/\phi_0)^3]$
close to its minimum and $V=V_\text{dS}$ away from it.
Because the self-interactions are polynomial, expectation values of observables 
are moments of the probability density $\vert\Psi_t^{(0)}\vert^{2}$.
It is convenient to add an auxiliary source term to the
generating functional
\begin{eqnarray}
\mathcal{Z}_t[J]
=	
\int_{\mathcal{C}(\Sigma_t)}	\mathcal{D}\phi \; 
\vert\Psi^{(0)}_t[\phi]\vert^{2} \exp{\left(\mathcal{J}_t[\phi]\right)}
\; ,
\end{eqnarray}
with $J := \left(1/\sqrt{\text{det}(q)(t)}\right) \, \delta \mathcal{J}_t[\phi]/\delta\phi$.
In closed form, 
\begin{eqnarray}
	\mathcal{Z}_t[J]
	=
	\vert\mathcal{N}^{(0)}_{t}\vert^2 \;
	\exp{\left(\tfrac{1}{2} \, [J]\mathcal{K}_t^{-1}[J] \right)}
	\; ,
\end{eqnarray}
where $\mathcal{K}_t^{-1}$ is the covariant functional inverse of $\mathcal{K}_t$. 
It is easy to solve $[f_1](K_t\star K^{-1}_t)[f_2] = [f_1]\delta[f_2]$ asymptotically. 
Close to the geodesic border
\begin{eqnarray}
	\label{Kinv}
	\mathcal{K}^{-1}_t
	\approx
	\text{i} \, t \text{ln}\left(t/t_\text{in}\right)
	\left[
		\delta + \frac{\mathcal{R}}{\text{ln}\left(t/t_\text{in}\right)}
	\right]
\end{eqnarray}
up to order $\text{ln}^{-2}(t/t_\text{in})$. 
Note that if the square bracket in (\ref{Kinv}) is replaced 
with an expression that is regular in the coincidence limit
then $\mathcal{K}_t^{-1}\rightarrow 0$ for $t/t_\text{in}\rightarrow 0$.
Close to the potential minimum and for $t\in]0,t_\text{in}]$
\begin{eqnarray}
	&&\langle\Psi^{(0)}_t | V(\Phi) |\Psi^{(0)}_t\rangle
	\approx
	\nonumber \\
	&&=V_\text{dS}
	\left(
	\frac{1}{\phi_0}\frac{1}{\sqrt{{\rm det}(q)}}\frac{\delta}{\delta J}\right)^2 
	\mathcal{Z}_t [0]
	\nonumber \\
	&&\approx
	(V_\text{dS}/\phi_0^2) \; \vert\mathcal{N}^{(0)}_{t_\text{in}}\vert^2
	\exp{(2\Gamma_t)} \; K_t^{-1}(0)
	\; ,
\end{eqnarray}
where $K_t^{-1}(0)$ denotes the regularized bilocal kernel function 
corresponding to $\mathcal{K}^{-1}_t$ in the coincidence limit. 
This can be used as an anchor for an inductive procedure proving 
that self-interactions have vanishing probabilistic support towards
the geodesic border of asymptotic Kasner geometries. 
Self-interactions respect the ground state. In this sense
Kasner geometries in the immediate vicinity of $\Sigma_0$ protect
the ground state against non-Gaussian deformations, and in turn,
the ground state renders these geometries quantum complete. 
This is a consequence of the Belinskii-Khalatnikov-Lifshitz conjecture 
which is at work here.  
At the operator level, effectively $\mathcal{H}\approx \mathcal{T}(\Pi;g)$
close to the geodesic border. 

\section{Discussion \& Conclusion}
The main result of this article is the following:
Kasner universes are quantum complete preludes to inflationary 
spacetimes that smoothly transits to ordinary Bianchi-type I cosmologies. 
They still have geodesic borders beyond which spacelike singularities 
are lurking. Nonetheless these singularities do not constitute a physical pathology as such. 
For these singularities to present a physical pathology, they must admit  
a description based on observables and measurement processes. 
It is not sufficient to evaluate observables at the geodesic border, because it is 
a priori not clear whether the fields composing observables and measurement
devices are supported at or close to the border. For an imperfect analogy: 
Newton's potential is singular in the coincidence limit. However, this is not sufficient 
to conclude that Newtonian gravity is plagued by physical pathologies. 
In order to promote the mathematical singularity to a physical pathology,
initial conditions are required to exist such that the coincidence limit
can dynamically be realized in a finite time. 
The same holds for spacetime singularities. 

Certainly, curvature invariants are singular when evaluated at the geodesic borders
of Bianchi type-I cosmologies. Again, the question is whether this evaluation corresponds
to a physical measurement process. 
As an imperfect analog consider the motion of a classical electric charge
in a repulsive Coulomb potential on the half line $[0,\infty)$. Clearly the 
Coulomb potential has a mathematical singularity at the origin. 
However, any electrical charge with finite initial energy moving towards the origin
will never be able to reach the origin, because it gets always reflected by the 
potential barrier. Therefore it can only probe the potential at an energy scale 
compatible with its initial energy. 
Perhaps a better analogy is given by a bound state electron within hydrogen.
An electron in a hydrogen bound state cannot probe the origin of the attractive Coulomb potential. 
Normalizability gives a vanishing probability for such an electron to be located at the origin. 
This has nothing to do with Heisenberg uncertainty, because uncertainty relations hold only for normalizable 
wave functions, and so normalizability has to be imposed first, which proves to be sufficient to guarantee
a vanishing probability measure at the origin.

Describing geodesic borders or their neighborhood
in terms of observables requires to evolve fields from an initial surface 
towards the border. The evolution is not governed by a one-parameter group of
isometric operators, but instead by a semigroup. As a consequence, norms
of quantum states are not conserved, whereby quantum states refer 
to wave functionals over the space of field configurations in the Schr\"odinger
representaton of quantum field theory. Semigroups appear here naturally since 
the spacetime is assumed to be absolutely inert against quantum fluctuations. 
Any friction-like phenomena caused by this background reduces the state norm. 
Albeit the dynamics are not given by a unitary evolution, a probabilistic interpretation 
of the wave functional is still possible. The situation is analogous to the treatment
of open quantum systems. This implies, however, that the spectrum of 
the Schr\"odinger Hamiltonian is not essentially self-adjoint. Of course, there is 
nothing wrong with this: Damping phenomena give rise to complex dispersion 
relations with a finite imaginary contribution. 
It is still possible to think of Hamiltonians as evolution generators 
in the usual infinitesimal sense. The pair (unitary, selfadjoint) is superseded 
by (contractive, accretive) in the spirit of Stone's theorem. This is the reason
why it is convenient to consider evolution semigroups. In fact, there is no alternative:
The geodesic borders considered here enforce a contractive evolution although
the quantum fields become free in their vicinity. Even in this situation, the Hamiltonian 
is not a symmetry generator simply because the evolution is considered in 
an approximate asymptotic Kasner spacetime.  

So Bianchi type-I cosmologies are quantum complete and geodesic incomplete. 
This apparent clash of completeness concepts requires a resolution. The resolution 
is straightforward: Singularity diagnostics is a well-established 
method directly related to the mathematical spacetime model. 
It is based on the geodesic motion of point particles. Its physical realization 
implies an event set described by classical point particle phenomenology.  
This phenomenology is invalid close to spacelike geodesic borders 
where absorption and emission processes are operative. 
Classical particle phenomenology gives the wrong description of the 
observable events close to the geodesic border, and therefore it 
gives the wrong description of the physical spacetime bordering
on spacelike singularities. An adequate characterization of the physical situation
at the border requires quantum field theory. Once this is established, statements
referring to Bianchi type-I cosmologies or their asymptotic Kasner approximations  
as geodesic incomplete are mathematically correct but physically irrelevant. 
Therefore it would be appropriate to call Bianchi type-I cosmologies just complete. 

We have shown that Bianchi type-I cosmologies with approximate Kasner 
geometries close to their geodesic border and inflationary spacetimes 
away from the borders are quantum complete realizations of inflation.
The argument is completely universal and holds for any inflaton potential.
This is a reflection of the famous conjecture by Belinskii, Khalatnikhov and
Lifshitz: In approximate Kasner regions, temporal changes dominate over any spatial correlations,
and hence the inflaton evolution becomes 
effectively free, which is why these types of geodesic borders allow
universal statements. 
\begin{acknowledgements}
It is a great pleasure to thank Abhay Ashtekar, Eugenio Bianchi, Kristina Giesel, Patrick Hager, Maximilian K\"ogler, Florian Niedermann, and Graham Shore for inspiring discussions. 
We appreciate financial support of our work
by the DFG cluster of excellence 'Origin and Structure of the Universe',
the TRR 33 'The Dark Universe' and the Alexander von Humboldt Foundation.
\end{acknowledgements}
\appendix

\section{Temporal gradients and Kasner geometries}
The conjecture of Belinskii, Khalatnikov and Lifshitz (BKL) stipulates 
that temporal gradients dominate over spatial gradients in the vicinity
of a spacelike singularity.   
We present a quick argument that this guarantees Kasner geometries
close to the geodesic border.

In a Gaussian normal neighborhood, $g=-\text{d}t\otimes\text{d}t + h$ and
the BKL conjecture amounts to requiring $|\partial_t h| \gg |\partial_x h|$.
The Levi-Civita connection is given by 
\begin{eqnarray}
	\Gamma^t_{ab} = \tfrac{1}{2} \partial_t h_{ab} + \mathcal{O}(\varepsilon)
	\; , \hspace{0.1cm}
	\Gamma^a_{bt} = \tfrac{1}{2} g^{ac} \partial_t h_{cb} + \mathcal{O}(\varepsilon)
	\; .
\end{eqnarray}
Here $a,b,c$ are spatial indices and all other Christoffel symbols are
$\mathcal{O}(\varepsilon)$, where $\varepsilon := |\partial_x h| / |\partial_t h|$
serves as the smallness parameter. Note that $\Gamma^t_{ab}$ are the
extrinsic curvature components $K_{ab}$ in this coordinate neighborhood. 
Therefore, up to $\mathcal{O}(\varepsilon)$, 
\begin{eqnarray}
	&&\left(\text{Ric}\right)_{tt}
	=
	-\partial_t \text{Tr}_h(K) - \text{Tr}_h\left(KK\right) \; ,
	\\
	&& \left(\text{Ric}\right)_{ab}
	=
	\partial_t K_{ab} - \left(K K\right)_{ab} \; ,
\end{eqnarray}
with $(KK)_{ab} := K_{ac} h^{cd} K_{db}$. 
Note that $(\text{Ric})_{ta}=\mathcal{O}(\varepsilon)$. This poses no problem
provided $h$ is diagonal since then $(\text{Ric})_{ta}=0$. 
Assuming $h$ is diagonal, $h_{ac} = \delta_{a\hat{c}} w_{\hat{c}}$, the extrinsic
curvature becomes 
$K_{ac} = \delta_{a\hat{c}} \dot{w}_{\hat{c}}/2$, where hats over spatial indices suspend 
the usual summation convention, and overdots denote time derivatives. For a vacuum solution
\begin{eqnarray}
	\left(\text{Ric}\right)_{tt}
	=
	\frac{1}{2}\sum_{a=1}^3 \left[\frac{\ddot{w}_a}{w_a} - \frac{1}{2}\left(\frac{\dot{w}_a}{w_a}\right)^2\right]
	= 0
\end{eqnarray}
This is solved by $w_a = t^{2p_a}$, provided 
$p_1+p_2+p_3 = p_1^{\; 2}+ p_2^{\; 2}+ p_3^{\; 2}$ which is precisely the Kasner condition.

\bibliography{Referenzen}
\end{document}